\documentclass[final,5p,times,twocolumn]{elsarticle}
\usepackage{graphicx}
\usepackage{amssymb}
\usepackage{lineno}
\usepackage{color}

\journal{arXiv}

\begin{document}

\newcommand{\red}{\textcolor{red}}
\newcommand{\blue}{\textcolor{blue}}
\newcommand{\etc}{\textit{etc.}}
\newcommand{\ie}{\textit{i.e.\,}}
\newcommand{\eg}{\textit{e.g.\,}}
\newcommand{\etal}{\textit{et al.\,}}
\newcommand{\roha}{Rohacell\textsuperscript{\textregistered} }
\newcommand{\Tedlar}{Tedlar\textsuperscript{\textregistered} }

\begin{frontmatter}

\title{Assessing the Feasibility of Interrogating Nuclear Waste Storage Silos using Cosmic-ray Muons}

\author[UniNa,INFNNa]{F.\,Ambrosino}
\author[INFNFi]{L.\,Bonechi}
\author[UniNa,INFNNa]{L.\,Cimmino}
\author[INFNFi,UniFi]{R.\,D'Alessandro}
\author[Glasgow]{D.\,G.\,Ireland}
\author[Glasgow]{R.\,B.\,Kaiser}
\author[Glasgow]{D.\,F.\,Mahon}
\author[CSFNSM,INFNFi]{N.\,Mori}
\author[INFNNa]{P.\,Noli}
\author[UniNa,INFNNa]{G.\,Saracino}
\author[NNL]{C.\,Shearer}
\author[INFNFi,UniFi]{L.\,Viliani}
\author[Glasgow]{G.\,Yang}

\address[UniNa]{Department of Physics, University of Naples Federico II, Naples , Italy}
\address[INFNNa]{INFN, sezione di Napoli, I-80126 Naples, Italy}
\address[INFNFi]{INFN, sezione di Firenze, I-50019 Sesto Fiorentino, Florence, Italy}
\address[UniFi]{Department of Physics and Astronomy, University of Florence, I-50019 Sesto Fiorentino, Florence, Italy}
\address[Glasgow]{SUPA, School Of Physics \& Astronomy, University of Glasgow, Kelvin Building, University Avenue, Glasgow, G12 8QQ, Scotland, UK}
\address[CSFNSM]{Centro Siciliano di Fisica Nucleare e Struttura della Materia (CSFNSM), I-95125 Catania, Italy}
\address[NNL]{National Nuclear Laboratory, Central Laboratory, Sellafield, Seascale, Cumbria, CA20 1PG, England, UK}

\begin{abstract}
Muon radiography is a fast growing field in applied scientific research.  In recent years, many detector technologies and imaging techniques using the Coulomb scattering and absorption properties of cosmic-ray muons have been developed for the non-destructive assay of various structures across a wide range of applications.   This work presents the first results that assess the feasibility of using muon radiography to interrogate waste silos within the UK Nuclear Industry.  Two such approaches, using different techniques that exploit each of these properties, have previously been published, and show promising results from both simulation and experimental data for the detection of shielded high-Z materials and density variations from volcanic assay.   Both detector systems used are based on scintillator and photomultiplier technologies.

Results from dedicated simulation studies using both these technologies and image reconstruction techniques are presented for an intermediate-sized legacy nuclear waste storage facility filled with concrete and an array of uranium samples.   Both results highlight the potential to identify uranium objects of varying thicknesses greater than 5\,cm within real-time durations of several weeks.  Increased contributions from Coulomb scattering within the concrete matrix of the structure hinder the ability of both approaches to resolve similar objects of 2\,cm dimensions even with increased statistics.  These results are all dependent on both the position of the objects within the facility and the locations of the detectors.  Results for differing thicknesses of concrete, which reflect the unknown composition of the structures under interrogation, are also presented alongside studies performed for a series of data collection durations.  It is anticipated that with further research, muon radiography in one, or both of these forms, will play a key role in future industrial applications within the UK Nuclear Industry.  

 \end{abstract}

\begin{keyword}
Muon Radiography \sep Scintillator Detectors \sep Nuclear Waste 
\PACS 96.50.S- \sep 29.40.Mc \sep 89.20.Bb
\end{keyword}
\end{frontmatter}

\section{Introduction}
When high-energy cosmic rays bombard the Earth's atmosphere, highly-penetrating showers of muons are produced and subsequently observed at sea level with a flux of approximately one muon per square centimetre per minute.    As charged particles, muons interact with matter primarily through ionising interactions with atomic electrons and via Coulomb scattering from nuclei.  Both of these mechanisms have been exploited in recent years within the field of muon radiography (alternatively, muography) to probe the internal composition of shielded and/or large structures that cannot be probed via conventional imaging techniques such as X-rays.   Since the pioneering experiment by E.\,P.\,George measured the thickness of the ice burden above the Guthega-Munyang tunnel in Australia in the 1950s~\cite{George1955}  and L.\,W.\,Alvarez conducted his search for hidden chambers in the Second Pyramid of Chephren in Egypt~\cite{Alvarez1970} a decade later, there has been a wealth of wide-ranging applications that have made use of cosmic-ray muons for imaging purposes, such as in volcanology~\cite{Tanaka2005,Ambrosi2011,Ambrosino2014}, nuclear contraband detection for national security~\cite{Borozdin2003a,Gnanvo08} and in the characterisation of legacy nuclear waste~\cite{Jonkmans2013,Clarkson2014a,Clarkson2014b,Clarkson2014c}.

Recent results presented in Refs.~\cite{Borozdin2012,Miyadera2013} first investigated the comparison between muon absorption (or transmission) and Coulomb scattering techniques for the assay of the tsunami-damaged reactors at the stricken Fukushima-Daiichi facility in Japan.  For this particular scale of application, the scattering approach provided better sensitivity to the core material and potential voids within several weeks of simulated exposure to cosmic-ray muons.  Subsequent research applied this Coulomb scattering technique to the experimental interrogation of the AGN-201M test reactor at the University of New Mexico~\cite{Perry2013} and the Toshiba Nuclear Critical Assembly reactor~\cite{Morris2014} to verify the potential of these image reconstruction techniques for an application of this nature and scale.


The results presented in the current work are the first from collaborative research performed by INFN (and the Universities of) Napoli \& Firenze and the Nuclear Physics group at the University of Glasgow to assess the feasibility of using these two complimentary techniques for the interrogation of nuclear waste storage silos.   In particular, the ability to identify small quantities of high-Z (atomic number) materials, in this case uranium, within the large, predominantly low-Z volume is investigated.  This volume of the structure is an `intermediate sized' $\mathcal{O}$(10m$^{3}$) application compared with the legacy waste containers $\mathcal{O}$(1m$^{3}$) and volcanoes  $\mathcal{O}$($>$100m$^{3}$) that the authors have previously shown could be successfully imaged using Coulomb scattering and absorption techniques respectively.  Hence, investigations using dedicated simulations are necessary to understand the capabilities and limitations of each technique and to therefore determine which is best suited to identify high-Z materials within a structure of this magnitude.  

The two differing approaches to image reconstruction are described in Section~\ref{sec:Approaches} with the corresponding detector technologies proposed for each study outlined in Section~\ref{sec:Technologies}.  The simulated detector and waste silo geometries are described in Section~\ref{sec:Simulation} with detailed analysis of the generated muon distribution used for both sets of studies.  Results showing the results from both techniques are presented and compared in Section~\ref{sec:Results} with implications for future industrial deployment discussed.

\section{Image Reconstruction Techniques for Muography}\label{sec:Approaches}
The two different approaches to image reconstruction using the Coulomb scattering properties and the energy loss of the muon are outlined, respectively, in the following subsections.

\subsection{Multiple Coulomb Scattering}
Seminal work outlined by Borozdin \etal in Ref.~\cite{Borozdin2003a} revealed the potential to locate and characterise high-Z materials within shielded containers using the Coulomb scattering properties of cosmic-ray muons.  This approach relies on precision reconstruction of the initial and scattered muon trajectories by pairs of tracking modules either side of the interrogation volume, which are subsequently used to infer the most probable scattering location within the container under interrogation and the associated scattering density, denoted $\lambda$.  This parameter is known to exhibit an inherent dependence on the atomic number of the scattering material~\cite{Schultz04} \ie larger scattering angles are typically observed for objects of larger Z values.   

Previous results reported by the current authors confirmed the potential of this imaging technique using a constructed prototype detector system developed at the University of Glasgow based on scintillating-fibre and multi-anode photomultiplier tube technologies~\cite{Clarkson2014a,Clarkson2014b}.  Here, a configuration of low (stainless steel) and high-Z (lead and uranium) materials of several centimetres in dimension were successfully located, with millimetre-precision, within a volume of air.  Clear discrimination between each material was observed.  Results published later in Ref.~\cite{Clarkson2014c} using the same detector setup demonstrated, for the first time, the potential of this technology and software in identifying and characterising high-Z materials encapsulated within a concrete-filled steel container.  The same uranium and lead samples were again imaged with millimetre-precision, albeit with suppressed $\lambda$ values compared with those reconstructed within an air matrix.  This can be attributed to the increase in Coulomb scattering from the concrete that acted to reduce the observed scattering effect from the high-Z material.

These results were all obtained using custom-developed image reconstruction software based on the MLEM algorithm introduced by Schultz \etal in Ref.~\cite{Schultz07}.   Prior to event-by-event data analysis using this software, the volume to be investigated is divided into small volume elements called voxels.  Studies have shown the size of these voxels to influence the achievable image resolution for a given amount of data~\cite{Clarkson2014b}.   However, for larger volumes as in this application, the minimum voxel size is constrained by the computing power available during image processing.  For the scattering results presented in this paper, cubic voxels of dimension 2.5\,cm were chosen unless otherwise stated.  For each muon, and for every voxel deemed to have been traversed, the scattering likelihood (and subsequently $\lambda$) is calculated in an iterative process and stored in the memory of that voxel.  Over the course of many muons, the median of the stored likelihood values in each voxel is used to determine the updated scattering density after each iteration.  Upon convergence, the most likely $\lambda$ value per voxel has been determined.  It is this parameter, quoted as a ratio of the average uranium and concrete background values, that is used to determine the imaging contrast parameter in these studies.



\subsection{Muon Absorption}\label{sub:MuonAttenuation}
The most common application of the muon absorption technique is the investigation of large structures. In particular, it is nowadays widely used in the investigation of geological structures like volcanoes. The measurement principle relies on the muon energy loss when interacting with matter: muons traversing high-density regions lose more energy and thus are more likely to be stopped before reaching the detector. The density of impinging muons coming from a certain direction across the target is thus a direct measurement of the column density of the target along that direction. In order to normalise the density map for the angular dependence of the impinging muon events, the measured density map is compared to a reference density map, usually obtained by pointing the detector at the open sky. This approach allows to obtain a two-dimensional projection of the investigated structure. 

The technique has been previously applied by other groups to the study of the damaged core of the Fukushima nuclear power plant and of the core of the Toshiba facility at Kawasaki. Real data and Monte Carlo simulations~\cite{Borozdin2012,Morris2014} show almost similar performance in detecting a high-density, high-Z core (\ie nuclear fuel) inside a containing structure depending on the geometry under study.

In this work the application of a similar technique to the nuclear waste storage scenario has been investigated. Results of Monte Carlo simulations show very promising performance in detecting small high-density nuclear fuel fragments, a few centimetres in size, placed inside large concrete silos, a few metres in size.

Detailed information of the track angle and the impact point on the detector is used to reconstruct the high-density object position after projecting the incoming tracks up to a vertical plane passing through the object itself. The density map of the reconstructed tracks is then subtracted from the map of the expected signal obtained from a simulated scenario where no high-density object is present inside the target silo. On the resulting map the signal shows up as a deficit of events coming from a certain sector on the map itself, thus identifying the high-density object. 

\section{Detector Technologies}\label{sec:Technologies}
The two radiographic techniques introduced in Section~\ref{sec:Approaches} rely on similar, though subtly different, detector technologies that were chosen to fulfil the unique requirements of each original application.  These existing systems are scaled up for the simulation studies presented in this work.  Both technologies make use of photomultiplier detectors and scintillator media, which will mitigate any issues arising from the employment of gas systems within the controlled environment.  These are proven to be robust, efficient and stable over prolonged periods of data collection.  These characteristics are essential for a potential operational system on a high-radiation, high-stress nuclear waste processing facility.  The two technologies are outlined in the following subsections.

\subsection{Scintillating Fibres \& Multi-anode Photomultipliers for Scattering Tomography}
Muon tomography relies on a detection medium with high spatial resolution to provide the precision measurement of small, milliradian-order scattering angles within the volume under interrogation \ie the assay volume.   To facilitate this, a small-scale prototype detector system based on Saint Gobain BCF-10 scintillating fibres was chosen with a fibre pitch of 0.2\,cm determined by simulation studies performed in Ref.~\cite{Clarkson2014b} to be optimal.  A modular detector design was fabricated with each module consisting of two orthogonal layers of fibres optically coupled to Hamamatsu H8500 Multi-Anode PhotoMultiplier Tubes (MAPMTs).  Two modules located above, and two below the assay volume are used to reconstruct the initial and scattered muon trajectories prior to image reconstruction. 

During extended data collection, the performance of a small-scale prototype system was continually monitored and found to operate stably with efficiencies in excess of 80\% per layer and with no measurable change in alignment.  With design improvements planned for a future full-scale system, efficiencies are expected to increase to over 95\% per layer with increased signal to noise separations also anticipated.  However, all timescales quoted in this work with regards to this system assume a 100\% muon detection efficiency for the (2\,m)$^{2}$ active area system simulated and shown in Section~\ref{sec:Simulation}.

\subsection{Scintillator Bars, WLS Fibres \& Silicon Photomultipliers for Radiography}
In this case, the required spatial resolution is not as demanding as in the Coulomb scattering scenario and the use of plastic scintillating bars coupled with Wave Length Shifting (WLS) fibres allows for the construction of large size detectors with contained costs. This kind of technology has been applied to the  MU-RAY experiment, devoted to the  study of volcanoes~\cite{Ambrosino2014}.  The bars have a hole along the axis, where a WLS fibre is placed. Since WLS fibres have optimal light transmission properties,  plastic scintillators, with unexceptional optical properties, can be used. In the case of the MU-RAY detector, bars with a length of 107\,cm have been used, while the same kind of bars, produced at FERMILAB-NICADD~\cite{FERMILAB} with a length up to  245\,cm have been used in  the MINERVA experiments~\cite{MINERVA}.  

Another advantage in the use of WLS fibres is that they can be easily coupled to Silicon PhotoMultipliers (SiPMs). SiPMs have better performances and less cost per channel with respect to standard or multi-anode phototubes.  Moreover SiPMs are robust and compact.  
 
Using scintillator bars with triangular cross section, it is possible to improve the spatial resolution, using a centre of charge technique~\cite{Anastasio2013}.  In the MU-RAY detector, triangular bars with a base of 3.3\,cm and a height of 1.7\,cm have been used.

\section{Geant4 Simulated Scenarios}\label{sec:Simulation}
The detection capability for high-Z materials of the two techniques have been studied by means of Monte Carlo simulations based on the GEANT4 toolkit~\cite{Allison2006}.

\subsection{Muon Event Generator}
Simulating a realistic muon flux is key to the studies performed in this paper. The spectral shape can affect the detection capabilities, with harder spectra being populated by more high-energy particles that are less likely scattered or absorbed, while the normalisation affects the estimation of the data acquisition time corresponding to a given number of Monte Carlo events. Phenomenological models of the muon flux at ground are available (\textit{e.g.} Ref.~\cite{Thompson1975}), as well as direct measurements (\textit{e.g.} Ref.~\cite{Hebbeker2002}) for various energy ranges but mostly for near-vertical incidence angles. 

For this study, the Monte Carlo muon generator has been modelled around the ground measurements reported in Ref.~\cite{Bonechi}. These measurements were performed with a magnetic spectrometer with an MDR of approximately 230\,GeV/c, placed at a geographic latitude of about 42\,degrees North, and pointing towards North. The range in momentum spans the interval 0.1\,-\,130\,GeV/c, while the polar angle range is 0\,-\,80\,degrees. Covering such wide ranges in momentum and polar angle with a single instrument is very important in order to minimise the effect of instrumental systematics. Moreover, this gives a consistent estimate of the vertical flux as well as the near-horizontal one, which is of particular importance in this study given the fact that both regimes are probed.

\begin{figure}[t] 
\centering 
\includegraphics[width=\columnwidth,keepaspectratio]{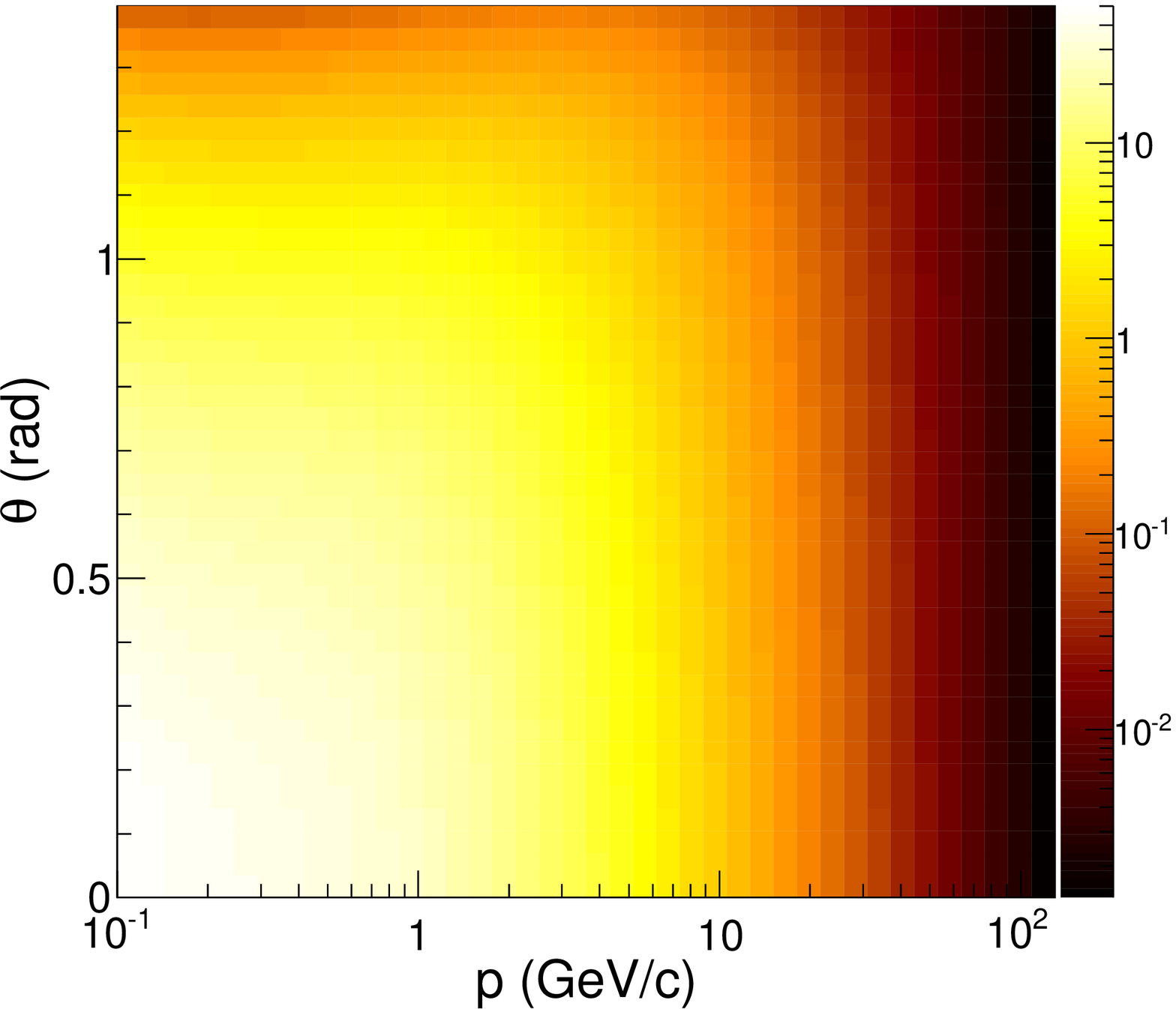}
\caption{Generated muon flux distribution as a function of polar angle $\theta$ and momentum $p$.  The colour indicates the differential flux detected by the ADAMO experiment~\cite{Bonechi} in units of muons\,m$^{-2}$\,sr$^{-1}$\,(GeV/c)$^{-1}$\,s$^{-1}$.}
\label{fig:BonechiFlux}
\end{figure}

The flux distribution implemented in the Monte Carlo muon generator is shown in Figure~\ref{fig:BonechiFlux}. The actual muon generation is performed by means of a Hit\&Miss algorithm that samples the initial momentum and polar angle according to this distribution, while the azimuthal angle and the initial position on the generation surface are sampled from uniform distributions.

\subsection{Waste Silo Geometry}

The implemented geometry of the storage silo consists of an external cylindrical shell made of reinforced concrete, with a height of 4\,m, external diameter of 4\,m and 0.5\,m thickness. A 1\,cm thick shielding made of stainless steel is present between the external shell and the internal part of the silo, which is filled with ordinary concrete. All the materials are considered to be homogeneous. Eventual inhomogeneities may result in systematic effects which could reduce the detection capability. This is true especially for the absorption technique in the case where the background map for a real, non-homogeneous silo is computed by means of Monte Carlo simulations of a homogeneous silo. These effects are not taken into account in this work.

Uranium samples of various dimensions have been placed at different locations within the internal part of the silo, in order to assess the dependence of the expected signal on size and position of the sample inside the silo. 

\subsection{Simulated Detector Positions}
\begin{figure}[t] 
\centering 
\includegraphics[width=\columnwidth,keepaspectratio]{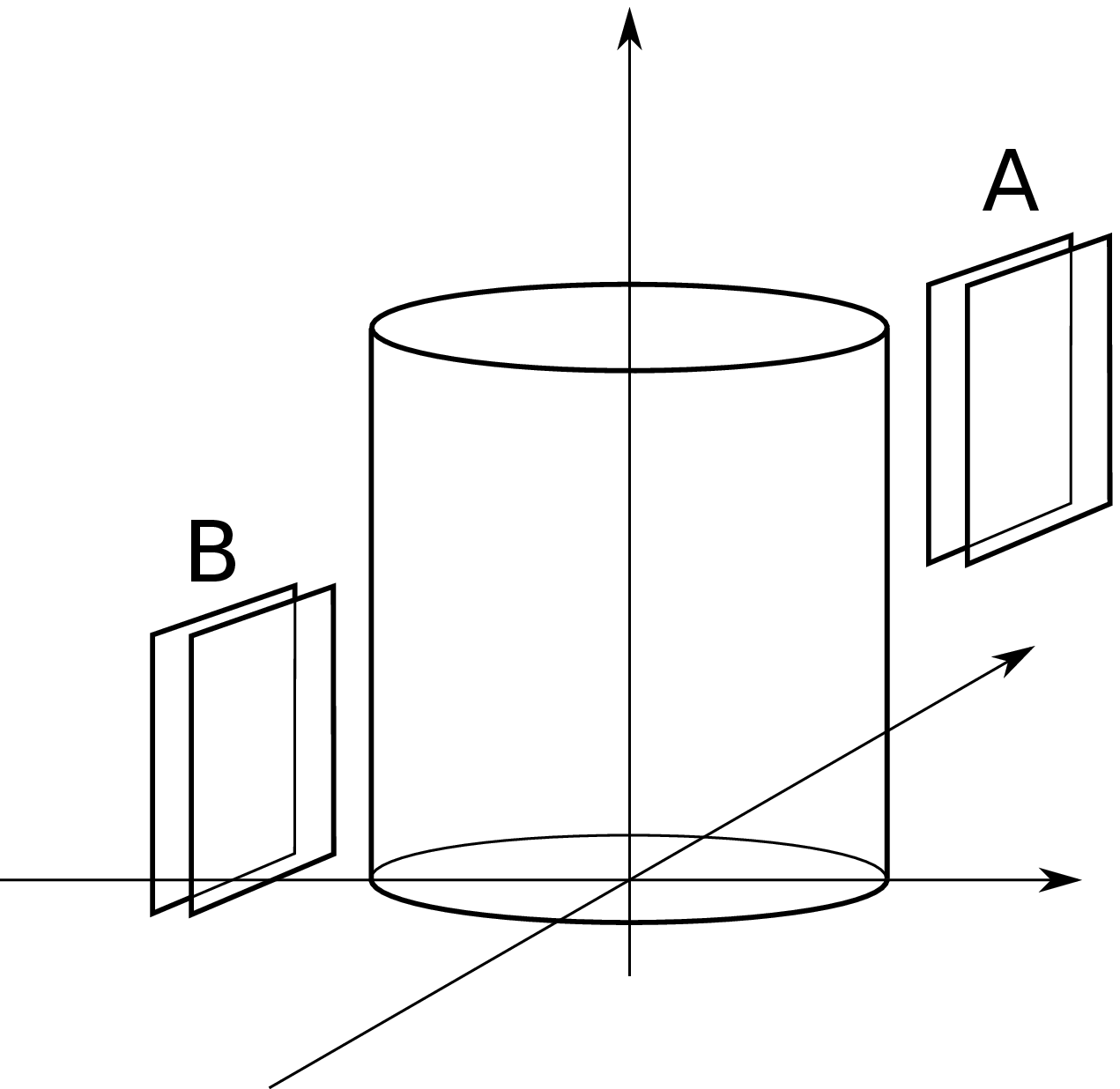}
\caption{Schematic representation of the positions of the detectors modules (each drawn as a pair of tracking layers) in Monte Carlo simulations with respect to a cylindrical storage silo. Both detectors pairs A and B are present in simulations for Coulomb scattering studies, while only detector pair B is used for absorption studies.}
\label{fig:SiloGeometry}
\end{figure}

The detector for absorption studies consisted of two 2\,m\,x\,2\,m scintillating layers made of vinyltoluene, placed 50\,cm apart from each other.  In order to speed up the simulation and simplify the reconstruction procedures, the layers were not segmented in elements, but implemented as a single volume. The impact point of the impinging muon is taken from the Monte Carlo truth, and a Gaussian smearing with $\sigma = 0.3$\,cm is applied in order to account for the finite resolution of a real device. The detector stands on the ground and its front face is 50\,cm from the silo as shown in Figure~\ref{fig:SiloGeometry}.

For the Coulomb scattering investigations, two orthogonal planes of 1000 scintillating fibres of 0.2\,cm pitch comprised each detection module in the simulation.  The bottom set of detector modules was placed at the same position as in the absorption studies and the top set of detector modules was positioned on the opposite side of the silo at the same distance of 50\,cm from the outer wall and centred on a height of 3\,m.   This placement is again illustrated in Figure~\ref{fig:SiloGeometry}.

\section{Simulation Results}\label{sec:Results} 
Results obtained from the two dedicated simulation analyses for the different image reconstruction approaches are detailed in the following subsections with comparison and discussion presented in Section~\ref{sub:Comparison}.

\subsection{Muon Tomography using Coulomb Scattering}\label{sub:GlasgowResults}

\begin{figure*}[t] 
\centering 
\includegraphics[width=2\columnwidth,keepaspectratio]{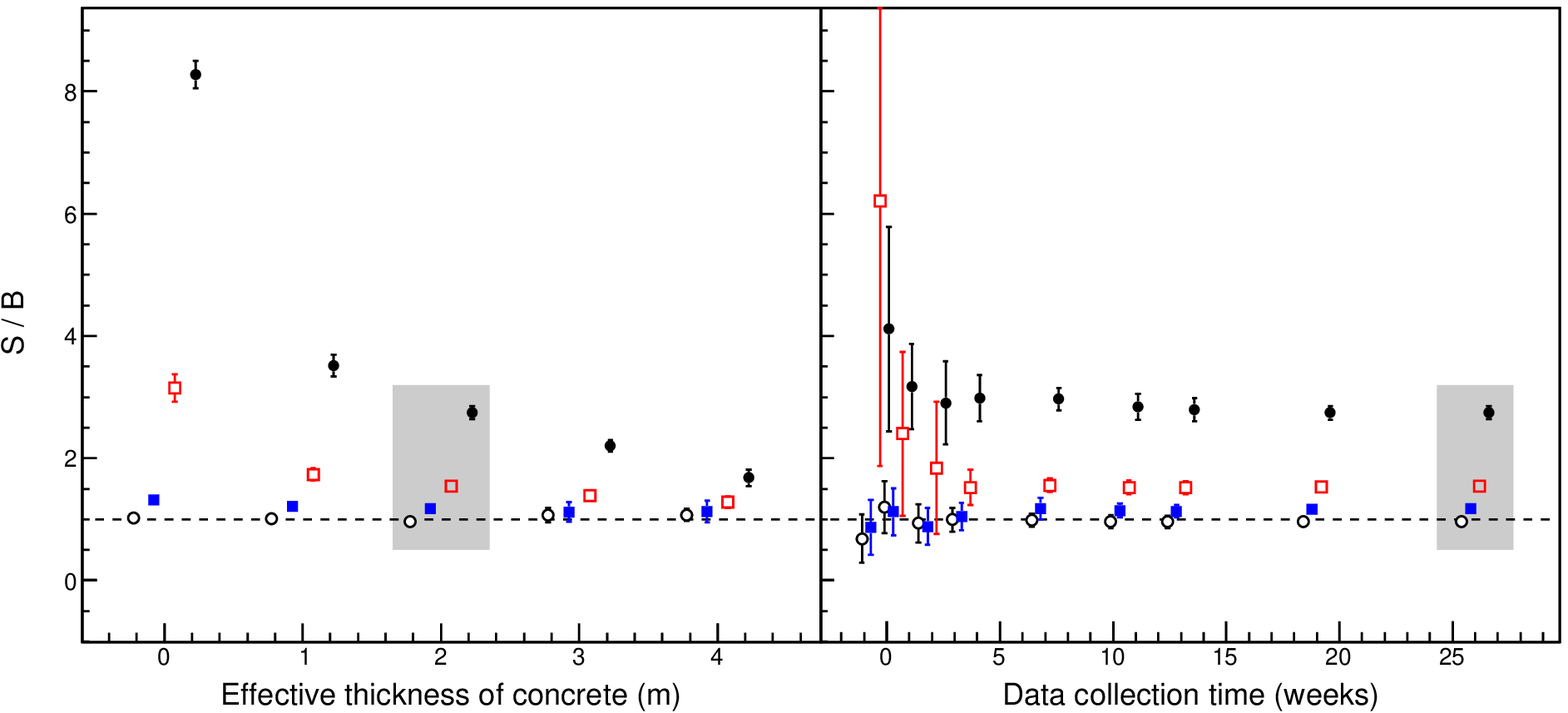}
\caption{The signal to background ratios (S/B) as a function of concrete thickness (left, for fixed collection time of 26\,weeks) and data collection time (right, for fixed concrete thickness of 2\,m) for cubic uranium blocks of 20\,cm (filled black circles), 10\,cm (open red squares), 5\,cm (filled blue squares) and 2\,cm (open black circles). Error bars represent the spread in $\lambda$ values in the uranium and concrete background regions for each cube.  The shaded boxes highlight the same data set for ease of comparison.  All timescales assume a detector efficiency of 100\% and do not account for any reduction in muon flux arising from neighbouring facilities on site.  These results are described in detail in the text.}  
\label{fig:MahonResult}
\end{figure*}

For the purposes of the studies presented in this subsection, four cubes of uranium (of dimensions 2\,cm, 5\,cm, 10\,cm and 20\,cm) were simulated in different planes in the central region of the silo.  Results obtained using the Coulomb scattering reconstruction method for these four uranium cubes are summarised in Figure~\ref{fig:MahonResult} for different effective thicknesses of concrete (Study 1) and for a range of data collection durations (Study 2).  In both of these studies, the simulated detector positions remained fixed as described in Section~\ref{sec:Simulation}.  Again, cubic voxels of (2.5\,cm)$^{3}$ were chosen for all studies shown here.  

For Study 1, five different concrete thicknesses ranging from 0\,m (\ie an air-only scenario) to 4\,m (\ie a fully concrete-filled silo) were investigated for approximately 26\,weeks of real-time data collection.  Here, an air-filled silo was represented by the 1\,m concreted case that took into account the silo wall thickness.  This choice of variable was motivated by the unknown composition and fill-levels of the storage structures on site.   For all effective thicknesses of concrete studied, the 20\,cm and 10\,cm uranium cubes were clearly identified with average $\lambda$ values above the level of the concrete background signal.   These values are presented in Figure~\ref{fig:MahonResult} as signal to background ratios, denoted S/B.  Here, the `signal' (`background') is defined as the mean $\lambda$ value from the region containing the uranium (concrete) sample.  

These S/B values ranged from 8.3 (3.2) for the air scenario, down to 1.7 (1.3) for the scenario containing a total concrete thickness of 4\,m for the 20\,cm (10\,cm) uranium cube.

With the same simulated muon exposure, the 5\,cm cube of uranium was discerned from the background for effective thicknesses of concrete of 2\,m and less, though with relatively small S/B values in the region of 1.2 \ie approximately 20\% above background.   For larger concrete thicknesses, \ie silos more than half full or those with a larger effective density, it was not possible to distinguish this volume of uranium.  In all instances, the 2\,cm cube could not be reconstructed using current scattering techniques.  This was likely the result of a combination of factors including the small angular acceptance imposed by the silo geometry, which effectively restricted the detection of larger scattering (and therefore, larger $\lambda$) events.   A further contributing source was the smallest achievable voxel size of (2.5\,cm)$^{3}$ for the whole volume dictated by computational constraints, which exceeded the size of the target object.  In this instance, the voxels containing the uranium sample would also contain significant contributions from concrete (or air depending on which thickness scenario was studied) that would act to dilute the effective scattering density reconstructed in that voxel.  However, studies were performed with (1.5\,cm)$^{3}$ voxels using a smaller assay volume and no discernible signal was observed in the region of the 2\,cm cube of uranium.  The most significant explanation for this was the large increase in Coulomb scattering introduced by the many metres of concrete between detector pairs that severely reduced the ability to resolve small high-Z objects.

\begin{figure}[t] 
\centering 
\includegraphics[width=\columnwidth,keepaspectratio]{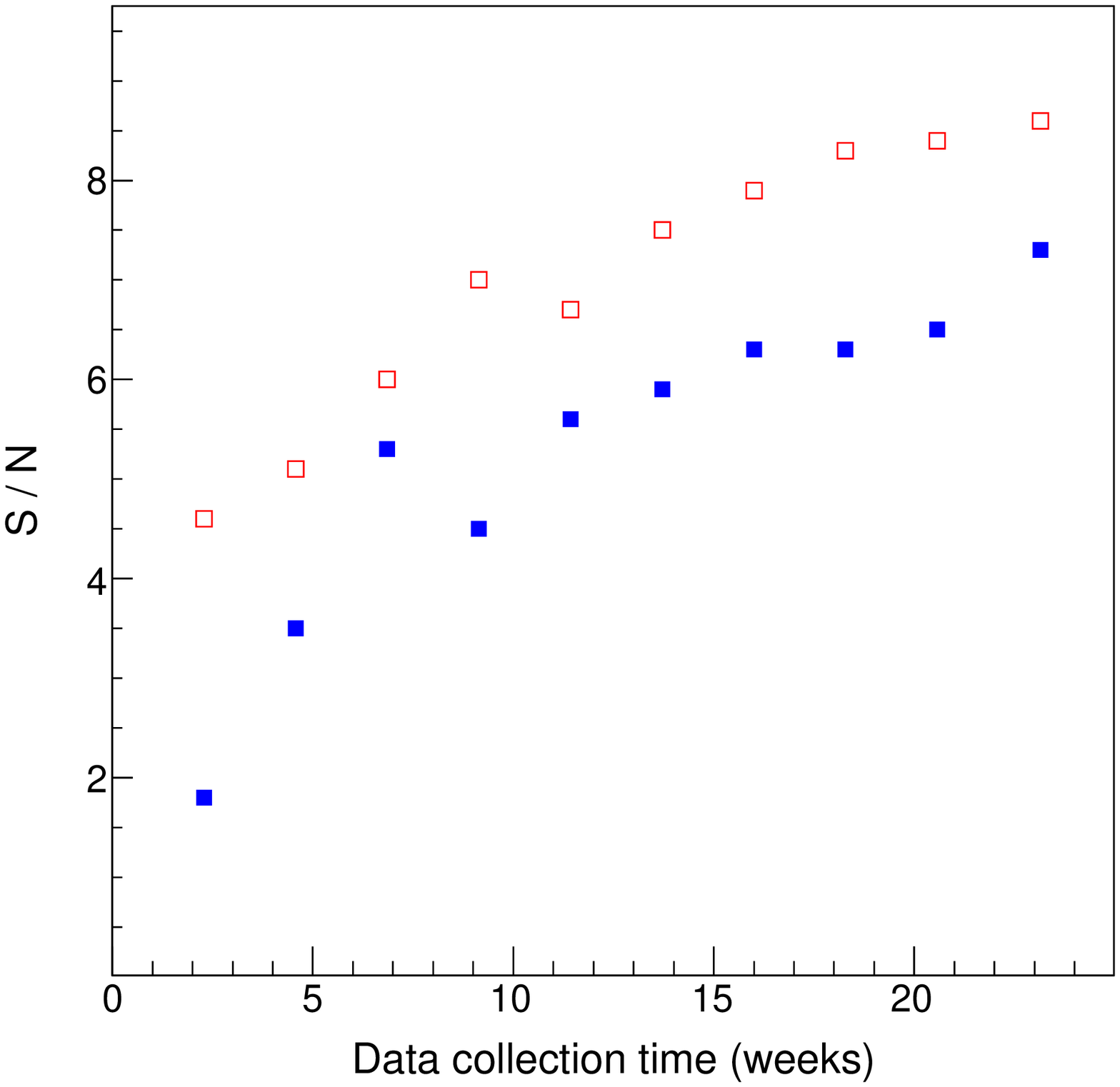}
\caption{The signal to noise ratios (S/N) as a function of data collection time for cubic uranium blocks of 10\,cm (open red squares) and 5\,cm (filled blue squares), placed at the center and at the top of the silo, respectively. These results are described in detail in the text.}  
\label{fig:MoriResult}
\end{figure}

The discriminating power of this technique (quantified as S/B) was also assessed as a function of data collection duration to provide an indication of the necessary timescale required to reliably identify samples of uranium of the studied dimensions.  Figure~\ref{fig:MahonResult} also shows these results for the 2\,m effective concrete thickness case for timescales ranging from one day up to 26\,weeks.   All quoted timescales were quoted for a 100\% detection efficiency and do not take into consideration any reduction in muon flux arising from neighbouring facilities on a typical nuclear waste processing site.  The largest of the high-Z blocks was clearly discerned across all timescales with, as to be expected, greater clarity with increased data.  No discernible improvement was observed beyond 20\,weeks.  The (10\,cm)$^{3}$ object was visible to some degree of reliability from one week of data.  However, a clear judgement on the presence of this object could only be made after 6\,weeks.   The two smallest cubes of uranium were more difficult to observe for the same reasons described previously.  However, the cube of 5\,cm dimension was observed above background after 20\,weeks with a S/B ratio of 1.2 with negligible uncertainty.  As with the previous study, it was not possible to reconstruct the smallest sample for the durations or concrete depths investigated.

\subsection{Muon Radiography using Absorption}
To investigate the performance of the absorption method, multiple uranium cubes of different sizes (2\,cm, 5\,cm, and 10\,cm) were placed inside the simulated concrete silo as before. The resulting signal on tomography layers  intersecting each sample were studied by means of density maps with (7\,cm)$^{2}$ bins. A total amount of data corresponding to more than 5\,months of data taking (assuming 100\% duty cycle and detection/reconstruction efficiency) have been simulated in order to assess the dependence of the signal to noise (S/N) ratio on acquisition time. In this study, the signal is obtained by means of a bin-by-bin subtraction of two two-dimensional maps, as explained in Section~\ref{sub:MuonAttenuation}. The value of a single bin of the signal map is thus obtained as the difference of two Poisson variables with mean $\mu_1$ and $\mu_2$. Approximating the values of $\mu_1$ and $\mu_2$ with the event counts $N_1$ and $N_2$ in the bins of the two maps, the S/N of each bin of the signal map is defined as:
\begin{equation}
S/N = \frac{N_1-N_2}{\sqrt{N_1+N_2}}
\end{equation}
 
The most relevant result is that there is a strong dependency of the detection capability on the size of the sample and its position. A 10\,cm size sample placed at the centre of the silo should be detectable with a data taking period of about 2\,months. The detection threshold here has been placed at S/N\,=\,5, given the fact that there is only a 0.08\% probability of having a S/N\,=\,5 due to a pure statistical fluctuation on the map bins corresponding to the silo area. A 5\,cm sample placed nearby was not visible at all. However, as anticipated, sample size is not the only relevant parameter for detection: the 10\,cm sample disappear when placing it in the bottom part of the storage silo, at a height of about 1\,m from the ground, and farther from the detector. This can be understood in terms of reduced statistics due to the lower intensity of near-horizontal flux and of the increased contribution of multiple scattering in the concrete between the uranium and the detector; this last effect can scatter muons so that they appear to come from the uranium region, diluting the missing events signal. Conversely, the 5\,cm cube becomes visible when placing it in the upper part of the silo (at a height of 3.5\,m) and nearer to the detector: the combined effect of increased statistics and reduced multiple scattering contribution make the sample visible with about 3\,months of data taking. The detailed dependence of the S/N on acquisition time is shown in Figure~\ref{fig:MoriResult} for the two described cases where the signal was measurable.

The signal from the 2\,cm cube has not been detected, regardless of its position inside the silo. Although the dimensions of the map bins may have a role in shaping the signal and determine the total S/N, no significant dependence of the detection capability for small samples on this parameter have been observed down to a minimum bin size of (2\,cm)$^{2}$. Multiple scattering in the concrete appears to be the limiting factor: turning off this process in the Monte Carlo simulation, the small 2\,cm sample was clearly visible, even with only a couple of months of statistics.

\subsection{Comparison of Results}\label{sub:Comparison}
The results presented in the previous subsections from the Coulomb scattering and absorption techniques both reveal the potential to identify and locate some, but not all, of the simulated uranium samples within the silo.   Though a direct comparison between the two techniques, and indeed the two conceptually-different imaging parameters S/B (scattering) and S/N (absorption), is not possible due to the nature of the two techniques, both were successful in reconstructing cubic samples of 5\,cm dimension and above.  Although this observation is subject to some caveats that are outlined in the following discussion.  

In principle, the reconstruction ability of these high-Z objects using the scattering technique should be largely independent of their location within the silo given a large detector area and angular coverage and/or optimised detector placement.   For the results shown in Figure~\ref{fig:MahonResult}, this is not the case with a proportion of muon events with large scattering angles (from high-Z materials) scattering outwith the limited angular acceptance.  The results from the chosen placements of uranium should however provide a reasonable estimate of the detection capabilities of high-Z objects of these sizes.  For the muon absorption results presented in Figure~\ref{fig:MoriResult}, the ability to resolve an object is strongly dependent on its location within the silo as outlined in the previous subsection.  Compared with the scattering scenario for the fully-concreted silo, there is the potential for objects with dimensions of around (5\,cm)$^{3}$ to be observed with far greater detection capability if located in more favourable regions of the silo.   However, the converse is also true in that a similarly-sized object in a less favourable location cannot be identified for the timescales investigated.  This observation highlights the complimentary nature of these two reconstruction techniques.

While the probabilistic S/N parameter provides a measure of the reliability of the observation (\eg a value of 5 indicates a 0.08\% probability of a statistical fluctuation), the S/B values determined from the reconstructed uranium and concrete $\lambda$ distributions do not provide a corresponding confidence metric.  However, the smoothness of these underlying distributions dictate the reliability of the determined S/B values with any statistical fluctuations taken into consideration within the quoted errors.

Both reconstruction techniques, at present, fail to resolve the (2\,cm)$^{3}$ piece of uranium as a direct result of the detrimental effect that the Coulomb scattering from the large volume of concrete has within the silo geometry.

\section{Summary} 
Imaging results have been presented from initial GEANT4 simulations of a small nuclear waste storage silo using two complimentary cosmic-ray muon radiographic techniques developed at INFN and the Universities of Napoli \& Firenze and the University of Glasgow.  Using the attenuating and Coulomb scattering characteristics of the muon respectively, uranium objects with dimensions of (10\,cm)$^{3}$ and greater were definitively resolved within the concreted structure in timescales in the region of one month, with initial indications as to their presence deducible within a shorter timescale.  For smaller objects of 5\,cm dimensions, the identification depends on the position within the silo and/or the effective thickness of the material in the path of the muon.  Samples of uranium smaller than this limit were not resolvable due to the large extent of Coulomb scattering from the thick concrete.  

With further research into the limitations of these two techniques for this application, radiographic techniques using cosmic-ray muons have the potential to play a key role in the interrogation of large-scale waste structures within the global nuclear industry in the future.


\section*{Acknowledgements}
The authors gratefully acknowledge the UK National Nuclear Laboratory and Sellafield Ltd., on behalf of the UK Nuclear Decommissioning Authority, for their funding of this initial investigation.


\end{document}